\documentclass[conference]{IEEEtran}
\IEEEoverridecommandlockouts

\usepackage{url}
\usepackage{amsmath,amsfonts}
\usepackage{algorithmic}
\usepackage{algorithm}
\usepackage{array}
\usepackage[caption=false,font=normalsize,labelfont=sf,textfont=sf]{subfig}
\usepackage{textcomp}
\usepackage{caption}
\usepackage{stfloats}
\usepackage{url}
\usepackage{verbatim}
\usepackage{graphicx}
\usepackage{cite}
\usepackage{float}
\usepackage{subcaption} 
\usepackage{listings}
\usepackage{xcolor} 
\usepackage{threeparttable}
\begin{document}
%

\title{An Evidence of Addressing Coherence Errors in VQE Observables by Pulse-level VQE Approach}

\author{\IEEEauthorblockN{1\textsuperscript{st} Xiaoan Lin}
\IEEEauthorblockA{Jilin University
Changchun, China 
linxa1319@ mails.jlu.edu.cn}}

\maketitle

\begin{abstract}
Quantum computing is an advanced area of computing that leverages the principles of quantum mechanics. Quantum computing holds the potential to revolutionize various fields by handling problems that are currently intractable for classical computers. This research focuses on Variational Quantum Eigensolvers (VQEs) in the Noisy Intermediate Scale Quantum (NISQ) era.
We introduce and evaluate over-rotation and under-rotation errors in the measurement process, which are critical for obtaining accurate expectation values of Hamiltonians. Our study aims to determine the extent to which these errors affect the estimated ground state energy and the computational cost in terms of optimization iterations. We conducted experiments on H$_2$ and HeH$^+$ molecules, varying the rotation angle, and recorded the estimated energy and optimization iterations.
Our findings indicate that the pulse-level VQE algorithm exhibits resilience to quantum errors in terms of accuracy. Additionally, our results suggest that less frequent calibration of measurement rotation pulses may be sufficient, thereby saving substantial time and computational resources. However, it is important to note that while accuracy remains stable, the iteration count may still vary, necessitating a trade-off between calibration frequency and computational cost.
Our work complements previous research by focusing on the observable aspect of VQEs, which has been largely overlooked. This detailed analysis contributes to a more comprehensive understanding of VQE performance in the NISQ era and supports the design and implementation of more efficient quantum algorithms.
\end{abstract}


\section{Introduction}
Quantum computing is an advanced area of computing that leverages the principles of quantum mechanics. Unlike classical computing, which uses bits as the smallest unit of data (where each bit is either a 0 or a 1), quantum computing uses quantum bits or qubits. These qubits can represent and store information in a way that allows them to be a superposition of 0 and 1. Quantum computing holds the potential to revolutionize various fields by handling problems that are currently intractable for classical computers. This includes drug discovery: Yudong Cao, J. Romero, and Alán Aspuru-Guzik discuss how quantum simulation can enable faster and more accurate characterizations of molecular systems compared to existing quantum chemistry methods\cite{cao2018potential}. The study by Pei-Hua Wang et al. provides an extensive review of various applications including protein structure prediction, molecular docking, quantum simulation, and quantitative structure-activity relationship (QSAR) models, all crucial for advancing drug discovery and development\cite{10082989}; optimization problems in logistics: Weinberg Jiang Chen et al. explore the use of the Quantum Approximate Optimization Algorithm (QAOA) for enhancing efficiency in manufacturing and supply chain management, demonstrating significant potential for quantum algorithms to improve operational performance\cite{Chen_Marcus_Leesburg_2021}. Benjamin C. B. Symons and colleagues provide a practitioner’s guide to quantum algorithms for optimization, discussing practical applications for noisy intermediate-scale quantum devices across various fields\cite{Symons_2023}; cryptography: Quantum cryptography is a recurrent theme where quantum computing is seen as a transformative tool for enhancing security. Quantum computing can provide new algorithms and techniques that can potentially break current encryption methods and develop unbreakable quantum cryptographic systems\cite{zahedinejad2017combinatorial, kretschmer2023quantum}; and even in the field of artificial intelligence: quantum computing is applied to solve complex computational problems that are inefficient with classical algorithms. This includes optimization in AI contexts such as machine learning, neural networks, and pattern recognition. The integration of quantum computing in AI aims to leverage quantum mechanical properties to enhance the speed and efficiency of learning algorithms\cite{9355449,liang2022variational, wang2022torchquantum, peng2024hybrid, ranjan2024proximl}. The current state of quantum computing is the noisy intermediate-scale quantum (NISQ) era, which was defined by John Preskill in 2018\cite{Preskill_2018}. In this era, quantum computers typically contain 50 to 1000 qubits with relatively low gate fidelity compared with classical computer. The noise comes from influence of environment, imprecision of control and measurement. And because of the limited number of qubits, implementation of quantum error correction is impossible which requires thousands of physical qubits to produce one logical qubit\cite{PhysRevA.86.032324, gambetta2017building}. But we can still make use of NISQ devices to do some practical applications by using Variational Quantum Algorithms (VQAs)  as VQAs can run on limited quantum circuit depth and have some resistance to quantum noise. VQAs use both classical and quantum computers with the classical computer performing optimization algorithm and the quantum computer determining the cost function (or its gradient)\cite{Cerezo_2021}. Variational Quantum Eigensolver\cite{Tilly_2022, fedorov2022vqe, tang2021qubit, kandala2017hardware, wang2023robuststate,liu2022layer} is a type of VQAs that computes the ground state energy of a Hamiltonian using the variational principle. Due to the noise, qubits have limited coherence times and error will accumulate as execution which limit the depth of the circuits. But the circuits need have enough parameters to export the Hilbert space of the system. Therefore, using the pulse parameters can give an advantage over gate parameters as the former way can provide more parameters in the same circuit depth~\cite{egger2023pulse,liang2024napa,liang2023spacepulse, cheng2023fidelity}. In this research, we focus on pulse level VQE which uses parameters of control pulses as the input of optimization. The primary problems we address in this research is the impact of quantum errors in the measurement steps on the performance of Variational Quantum Eigensolvers (VQEs) within the Noisy Intermediate Scale Quantum (NISQ) era. Specifically, we focus on the following aspects:
\begin{itemize}
    \item \textbf{Impact on Estimated Energy}: Despite the inherent resilience of VQEs to quantum noise within variational circuits, measurement noise can still significantly affect the final computed values. This research aims to understand the extent to which errors in the measurement process, such as over-rotation and under-rotation errors, influence the accuracy of the estimated ground state energy.
    \item \textbf{Impact on Optimization Process}: Another crucial aspect of our research is to evaluate how these measurement errors affect the optimization process. We investigate whether these errors increase the number of iterations required by the optimization algorithm to converge, thus potentially increasing the overall computational cost.
    \item \textbf{Practical Implications for Pulse Calibration}: Given that NISQ devices are highly susceptible to environmental influences, leading to frequent miscalibration of control pulses, our research also explores the practical implications of these errors. By simulating inaccuracies in the control pulses and evaluating their impact, we aim to determine whether frequent calibration is necessary, or if the system can maintain accuracy within acceptable limits despite these errors.
\end{itemize}
The paper is organized as follows:
In section I and II, we give some introduction and background knowledge to our research. Next, we discusses the importance of considering observable and talk about how to simulate those noise in measurement. Then we present and evaluate our experiment result in section V, finally, we discuss some related work and make conclusion in the end.

\section{Background}
\subsection{Qubit and Quantum Computing}
A qubit is a two-level system, which can be represented using Dirac notation as:
\[
|\psi\rangle = \alpha|0\rangle + \beta|1\rangle
\]
where \(\alpha\) and \(\beta\) are complex numbers. Furthermore, to ensure probability normalization, the following normalization condition must be satisfied:
\[
|\alpha|^2 + |\beta|^2 = 1
\]

\subsection{Systems of Multiple Qubits}
For a system composed of multiple qubits, the overall state vector is the tensor product of the state vectors of each qubit. For a system of n qubits, the state vector can be represented as:
\[
|\Psi\rangle = |\psi_1\rangle \otimes |\psi_2\rangle \otimes \cdots \otimes |\psi_n\rangle
\]
where each \(|\psi_i\rangle\) represents the state vector of the ith qubit.

\subsection{Evolution of the System}
The evolution of a quantum system is described by the Hamiltonian \(H\), and the time evolution of the state vector is given by the following unitary transformation:
\[
|\psi(t)\rangle = e^{-iHt/\hbar}|\psi(0)\rangle
\]
Here, \(e^{-iHt/\hbar}\) is a unitary matrix that transforms the system from its initial state \(|\psi(0)\rangle\) to its state at time t \(|\psi(t)\rangle\).
When the Hamiltonian \(H\) explicitly depends on time, i.e., \(H = H(t)\), the evolution of a quantum system cannot be simply described by the exponential \(e^{-iHt/\hbar}\). Instead, we use the time-ordered exponential given by:
\[
|\psi(t)\rangle = \mathcal{T} \exp\left(-i \int_0^t H(\tau) \, d\tau / \hbar\right) |\psi(0)\rangle
\]
where \(\mathcal{T}\) denotes the time-ordering operator that ensures operators are applied in chronological order.

\subsection{Dyson Series}
The time-ordered exponential can be expanded into a Dyson series, especially useful for handling small changes in the Hamiltonian or for short-time evolutions. The Dyson series is expressed as follows:
\begin{multline}
\mathcal{T} \exp\left(-i \int_0^t \frac{H(\tau)}{\hbar} \, d\tau \right) 
=  I - \frac{i}{\hbar} \int_0^t H(\tau) \, d\tau  \\+ \frac{(-i)^2}{2!\hbar^2} \int_0^t \int_0^{\tau_2} H(\tau_1) H(\tau_2) \, d\tau_1 \, d\tau_2 \quad + \cdots
\end{multline}

where:
- \(I\) is the identity operator.
- The first term involves a simple integral of the Hamiltonian.
- Each subsequent term involves higher-order integrals, with time-ordering to ensure proper sequence of application.

\subsection{Gate based compilation work flow}
Quantum programs are written in Quantum Programming Languages such as IBM's Qiskit\cite{gadi_aleksandrowicz_2019_2562111} and Googles's Cirq\cite{cirq_developers_2023_10247207}. They are abstract high-level languages which have sophisticated control flow and are device-agnostic. The programs are then compiled into Quantum Assembly Languages. After this compilation programs only include 1 or 2 qubit gates and can be considered as quantum circuit. Because Quantum Assembly Languages are still device-agnostic. The next step is to express the gates using basis gates that are directly supported by hardware. The final step is to generate the Pulse Schedule using look-up table of basis gates and control pulses. Control pulses are usually electrical signals that directly control the state of qubit.

\begin{figure} 
    \includegraphics[width=\linewidth]{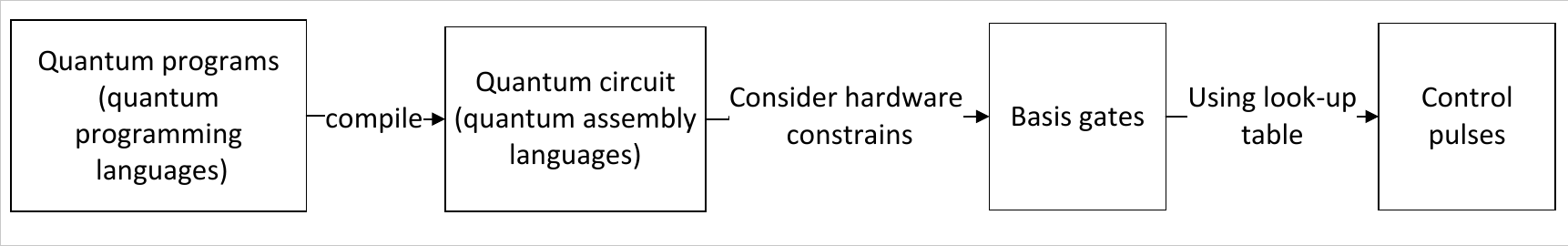} 
    \caption{Gate based compilation workflow.}
    \label{fig_1}
\end{figure}

\subsection{Variational Quantum Eigensolve}
For any state vector \(|\psi\rangle\) in the Hilbert space, we have the ground state energy following the condition
\begin{equation}
E_g \leq \langle \psi | H | \psi \rangle
\end{equation}
Define \(U(\Theta)\) as the evolution operator of the qubit system, where \(\Theta\) is a parameter determined by the control pulse schedule of the quantum computer. Then
\begin{equation}
E_g \leq \langle \psi(0) | U^\dagger(\Theta) H U(\Theta) | \psi(0) \rangle = C
\end{equation}
\(C\) is the cost function to be minimized in the optimization step.
The Hamiltonian, for which we want to find the ground state energy, can be written as a weighted sum of tensor products of Pauli operators
\begin{equation}
\hat{P}_\alpha = \bigotimes_{i=1}^n \sigma_{\alpha_i}, \quad \alpha_i \in \{ I, X, Y, Z \}
\end{equation}
\begin{equation}
H = \sum_\alpha \omega_\alpha \hat{P}_\alpha
\end{equation}
After optimization, we have
\begin{equation}
\Theta = \Theta_{\text{min}}
\end{equation}
\begin{equation}
E_g \leq C_{\text{min}} = \sum_\alpha \omega_\alpha \langle \psi(0) | U^\dagger(\Theta_{\text{min}}) \hat{P}_\alpha U(\Theta_{\text{min}}) | \psi(0) \rangle
\end{equation}
We can use \(C_{\text{min}}\) as the approximate ground state energy of the Hamiltonian.

\section{Motivation}

Variational Quantum Eigensolvers (VQEs) are known for their resilience to quantum noise, particularly within the context of variational quantum circuits. This resilience stems from the fact that, even under the influence of quantum noise, the system's evolution remains confined within its Hilbert space. Therefore, as long as the variational circuit has the capability to explore the entire Hilbert space, the presence of noise does not alter the final result. However, noise occurring during the measurement step can potentially affect the final outcome. To obtain the final result, it is necessary to measure the values of tensor products of Pauli operators:

\begin{equation}
\hat{P}_\alpha = \bigotimes_{i=1}^n \sigma_{\alpha_i}, \quad \alpha_i \in \{ I, X, Y, Z \}
\end{equation}

Due to hardware constraints, we typically can only measure the value of the Pauli Z operator. Consequently, before measurement, we need to apply rotations to the qubit states to obtain the values of the Pauli X and Y operators. This process can be mathematically described as follows.

Assume the initial state of a single qubit is a superposition state \(|\psi\rangle = A|+\rangle + B|-\rangle\), where \(|+\rangle\) and \(|-\rangle\) are the eigenstates of the Pauli X operator with eigenvalues +1 and -1, respectively. To measure the Pauli X operator, we apply a Hadamard gate \(H\) to rotate the state from the X-basis to the Z-basis. The Hadamard gate is defined as:

\begin{equation}
H = \frac{1}{\sqrt{2}} \begin{pmatrix} 1 & 1 \\ 1 & -1 \end{pmatrix}
\end{equation}

When we apply the Hadamard gate to the state \(|\psi\rangle\), the new state \(|\psi'\rangle\) becomes:

\begin{equation}
|\psi'\rangle = H|\psi\rangle = B|1\rangle + A|0\rangle
\end{equation}

Here, \(|1\rangle\) and \(|0\rangle\) are the eigenstates of the Pauli Z operator. By measuring this state in the Z-basis, we effectively measure the original state in the X-basis.

Similarly, to measure the Pauli Y operator, we apply a combination of gates: a \(\pi/2\) rotation around the X-axis (denoted as \(R_X(\pi/2)\)) followed by a Hadamard gate. The rotation matrix \(R_X(\theta)\) is given by:

\begin{equation}
R_X(\theta) = \begin{pmatrix} \cos(\theta/2) & -i\sin(\theta/2) \\ -i\sin(\theta/2) & \cos(\theta/2) \end{pmatrix}
\end{equation}

For \(\theta = \pi/2\), we have:

\begin{equation}
R_X(\pi/2) = \frac{1}{\sqrt{2}} \begin{pmatrix} 1 & -i \\ -i & 1 \end{pmatrix}
\end{equation}

Applying \(R_X(\pi/2)\) and then the Hadamard gate to the initial state \(|\psi\rangle\), we get:

\begin{equation}
|\psi''\rangle = H R_X(\pi/2) |\psi\rangle
\end{equation}

By measuring this state in the Z-basis, we can obtain the value of the Pauli Y operator.
Considering the rotations applied before measurement is crucial for several reasons:
\begin{itemize}
    \item \textbf{Performance Impact}: Noise in the measurement of observables can can lead to inaccuracies in the final computed values.
    \item \textbf{Optimization Iteration}: Even if the variational circuit can mitigate such errors, they may still affect the optimization process by potentially increasing the number of iterations required to reach convergence. This may result in an increased number of optimization iterations and significantly increases the computational cost.
\end{itemize}
In the current stage of quantum computing, noise is inevitable. Even if all control pulses are initially well-calibrated, they can become miscalibrated over time due to shifts in the system's Hamiltonian, and maintaining well-calibrated control pulses is extremely costly. 
Therefore, in this research, we aim to:
\begin{itemize}
    \item Evaluate the extent to which errors arising from overrotation and underrotation during the measurement process can influence the accuracy of the estimated energy provided by VQE.
    \item Determine how these errors impact the number of optimization iterations required, and thus the overall computational cost.
\end{itemize}
If those influence is found to be negligible, frequent calibration of the pulses may not be necessary.
Given these considerations, this issue is critically important and worth exploring in detail.
\section{Method}
\begin{figure*}[!t]
\centering
\includegraphics[width=\textwidth]{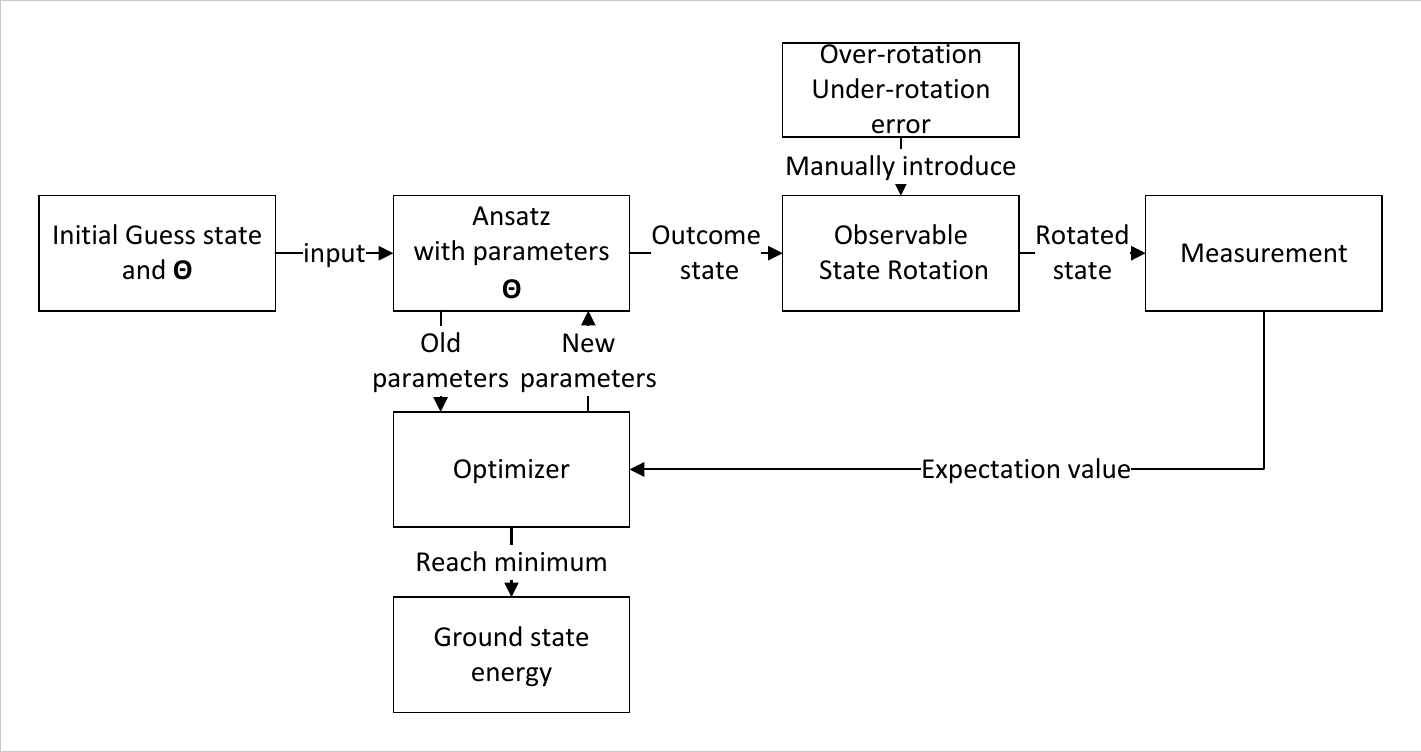}
\caption{Overview of workflow of our proposed work. We manually introduce over-rotation and under-rotation errors during the basis rotation process to simulate inaccuracies in control pulses. This allows us to investigate to what extent the ansatz can mitigate these errors and whether errors in the control pulses can increase the computational cost. }
\end{figure*}
\begin{figure*}[!t]
\centering
\includegraphics[width=\textwidth]{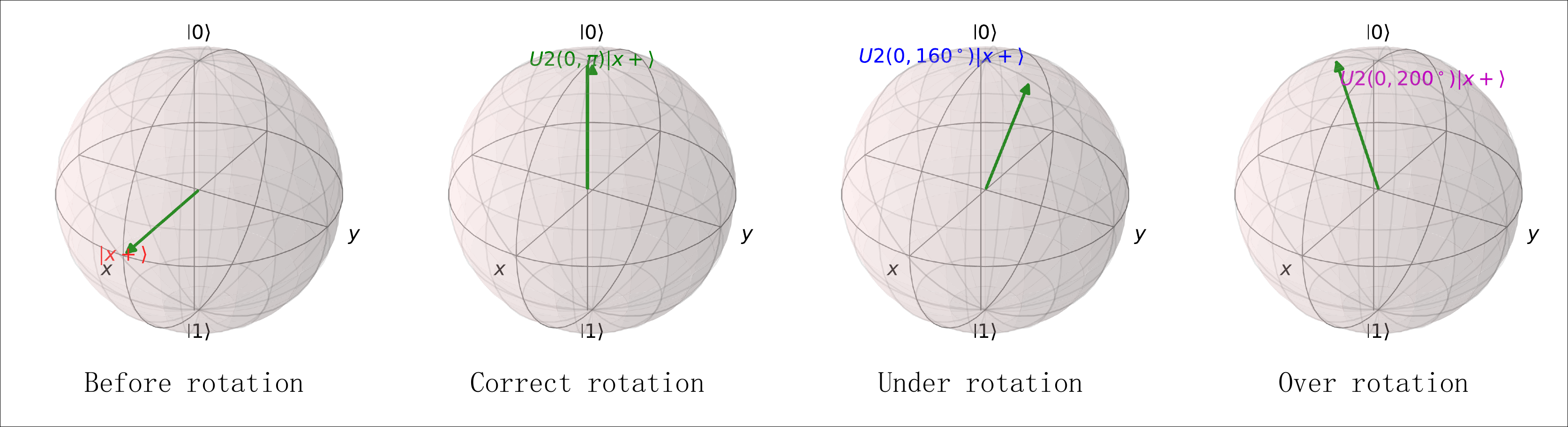}
\caption{This figure visualizes the process of basis rotation and explains the over-rotation and under-rotation errors on the Bloch sphere.}
\end{figure*}
\subsection{Pulse-level Vqe}
Using parameterized control pulses to construct ansatz may offer advantages in the following aspects (for detailed discussion see \cite{liang2023towards}):
\begin{itemize}
    \item Entanglement Capability:
    Due to hardware restrictions, the ability to generate highly entangled states is crucial for variational quantum algorithms. This capability can make the process of exploring the state space more efficient, thus improving the performance of VQAs.
    \item Parameter Dimension: 
    Parameter dimension measures the total number of independent parameters a quantum state defined by the parameterized quantum circuit (PQC) can express \cite{Haug_2021}.
    \item Sequence Duration: 
    Sequence duration is the execution time of a quantum program. Given the limited decoherence time of hardware, sequence duration is a critical factor to consider when designing circuits.
    \item Expressivity:  Expressivity refers to a circuit's capacity to generate pure states.\cite{Sim_2019}    
\end{itemize}
\subsection{U2 pulse}
To measure the expectation value of Pauli X, we need to apply a Hadamard gate before measuring the qubit. At the pulse level, this is done by applying a U2 pulse.

\[ 
U2(\phi, \lambda) = \frac{1}{\sqrt{2}}
\begin{pmatrix}
1 & -e^{i\lambda} \\
e^{i\phi} & e^{i(\phi + \lambda)}
\end{pmatrix}
\]

\begin{equation}
\begin{split}
U2(\phi, \lambda) &= e^{i\frac{\phi + \lambda}{2}} RZ(\phi) RY\left(\frac{\pi}{2}\right) RZ(\lambda) \\
&= e^{-i\frac{\pi}{4}} P\left(\frac{\pi}{2} + \phi\right) \sqrt{X} P\left(\lambda - \frac{\pi}{2}\right)
\end{split}
\end{equation}
if we set \(\phi\) = 0, and \(\lambda\) = \(\pi\) then
\[ 
U2(0, \pi) = H 
\]
Since we measure the expectation value of Pauli Z, the value of \(\phi\) will not influence the result. Therefore, only the error in \(\lambda\) matters. If \(\lambda > \pi\), it is over-rotation; if \(\lambda < \pi\), it is under-rotation, both of which may influence the final result.
\subsection{Introduce Error to the Observable Layer}
\begin{lstlisting}[language=Python, caption=Python Code for Introduce Error in the Observable.]
def measurement_pauli(prepulse, pauli_string, backend, n_qubit):
    with pulse.build(backend) as pulse_measure:
        pulse.call(copy.deepcopy(prepulse))
        for ind,pauli in enumerate(pauli_string):
            if(pauli=='X'):
                pulse.u2(0, np.pi+(np.pi/180)*(N), ind) #we introduce noise here.
            if(pauli=='Y'):
                pulse.u2(0, np.pi/2, ind)
        for qubit in range(n_qubit):
            pulse.barrier(qubit)
        pulse.measure(range(n_qubit))
    return pulse_measure
\end{lstlisting}
Before measure expectation value of Pauli X, we need to apply a H gate, in pulse level, this is done by using the U2 pulse. But due to shift of Hamiltonian of the system and other influence, the control pulse become inaccurate. So we want to simulate those errors and evaluate their influence of final result and computational cost.
We change the number \( N \) in the code to simulate over-rotation ( \(N > 0\) )and under-rotation ( \(N < 0\) ) errors and record the final outcome in each iteration.
Here is detailed explain of the code
\begin{itemize}
    \item prepulse : 
    prepulse is a pulse schedule generated by parameterized pulse circuit. Parameterized pulse circuit is the trainable layer of VQE. This layer is used to generate trail state \(U(\Theta) | \psi(0) \rangle\).
    \item pauli String : 
    For the task of simulating chemical molecules, we first use quantum chemistry software (such as Gaussian or PySCF) to perform electronic structure calculations and obtain the fermionic terms of the Hamiltonian. This involves selecting appropriate Gaussian-type orbitals (GTOs) as basis functions (e.g., STO-3G, 6-31G) to represent the molecular orbitals. Through self-consistent field (SCF) methods or other electronic structure calculation methods, we obtain the one-electron and two-electron integrals. The one-electron integrals \(h_{pq}\) and two-electron integrals \(h_{pqrs}\) are defined as follows:

\[
h_{pq} = \int \phi_p(\mathbf{r}) \left( -\frac{1}{2} \nabla^2 + V(\mathbf{r}) \right) \phi_q(\mathbf{r}) \, d\mathbf{r}
\]
\[
h_{pqrs} = \int \phi_p(\mathbf{r}_1) \phi_q(\mathbf{r}_2) \frac{1}{|\mathbf{r}_1 - \mathbf{r}_2|} \phi_r(\mathbf{r}_1) \phi_s(\mathbf{r}_2) \, d\mathbf{r}_1 \, d\mathbf{r}_2
\]

Here, \(\phi_p(\mathbf{r})\) are the chosen Gaussian basis functions and \(V(\mathbf{r})\) is the electron-nuclear potential.
These integrals are then used to construct the fermionic Hamiltonian of the molecule:
\[
\hat{H} = \sum_{pq} h_{pq} a_p^\dagger a_q + \frac{1}{2} \sum_{pqrs} h_{pqrs} a_p^\dagger a_q^\dagger a_r a_s
\]
where \(a_p^\dagger\) and \(a_q\) are fermionic creation and annihilation operators.

Next, we map these fermionic terms to Pauli strings using the Jordan-Wigner or Bravyi-Kitaev transformations. This process converts the fermionic operators into a linear combination of Pauli operators (e.g., XX, YY, ZI, IX, etc.). Specifically, the Jordan-Wigner transformation is applied as follows:
\[
a_p^\dagger = \frac{1}{2} (X_p - iY_p) \prod_{j=0}^{p-1} Z_j
\]
\[
a_p = \frac{1}{2} (X_p + iY_p) \prod_{j=0}^{p-1} Z_j
\]
By applying these transformations, we convert the fermionic Hamiltonian into a qubit Hamiltonian represented by a sum of Pauli strings:
\begin{equation}
\hat{P}_\alpha = \bigotimes_{i=1}^n \sigma_{\alpha_i}, \quad \alpha_i \in \{ I, X, Y, Z \}
\end{equation}
\begin{equation}
H = \sum_\alpha \omega_\alpha \hat{P}_\alpha
\end{equation}\cite{Tilly_2022}
    \item backend: 
    In our experiments, we utilize the \texttt{qiskit\_dynamics} backend, which is a specialized simulator provided by Qiskit for simulating open and closed quantum systems. This backend is particularly suited for modeling the dynamics of quantum systems under various conditions, including the presence of noise and environmental interactions.
The \texttt{qiskit\_dynamics} backend allows us to define the Hamiltonian of the quantum system, which dictates the energy levels and evolution of the system. Additionally, it supports the inclusion of Lindblad operators, which are essential for describing the dissipative processes in open quantum systems. This feature enables us to simulate more realistic scenarios where the quantum system interacts with its surroundings.
By leveraging \texttt{qiskit\_dynamics}, we can perform high-fidelity time evolution simulations, which are crucial for understanding the behavior of quantum circuits and the impact of various types of errors. This capability is essential for our study as it provides a comprehensive environment to test and validate our theoretical models and hypotheses under realistic conditions.
\end{itemize}
\section{Evaluation}

\subsection{Experiment Setting}
In this subsection, we describe the experimental setup.
\begin{itemize}
\item \textbf{QPulse}: In our experiments, we use the QPulse~\cite{liang2023towards} library, a benchmarking library that helps us evaluate the performance of Pulse VQE. QPulse provides a set of tools to generate, optimize, and analyze pulse sequences for quantum computing, enabling effective evaluation of quantum circuits at the pulse level.
\item \textbf{Qiskit Dynamics}: Qiskit Dynamics~\cite{puzzuoli2023algorithms} is our primary simulator for Pulse VQE. Given the limited accessibility of real quantum machines with pulse-level access, we use high-fidelity simulations of quantum systems to obtain reasonable results. Qiskit Dynamics supports the simulation of open and closed quantum systems, allowing the definition of Hamiltonians and Lindblad operators for high-precision time evolution simulations.
\item \textbf{Computational Environment}: Our simulations are performed on a high-performance computing server with the following configuration:
    \begin{itemize}
        \item \textbf{CPU}: 32 vCPU Intel(R) Xeon(R) Platinum 8352V CPU @ 2.10GHz
        \item \textbf{Memory}: 120 GB
        \item \textbf{Operating System}: Ubuntu 20.04 LTS
    \end{itemize}
This high-performance computing environment ensures that we can complete complex quantum simulations within a reasonable time frame and obtain high-precision simulation results.
\item \textbf{Optimizer Configuration}: In our experiment, we configure the optimizer with the following parameters:
    \begin{itemize}
        \item \textbf{Optimizer Name}: The name of the non-gradient optimizer used in the experiment. By default, we use COBYLA (Constrained Optimization BY Linear Approximations), which is a popular optimizer for quantum variational algorithms due to its efficiency in handling constrained optimization problems without requiring gradient information.
        \item \textbf{Maximum Iterations}: 100
        \item \textbf{Number of Shots}: The number of shots used for measurement in each iteration is 1024.
        \item \textbf{Rhobeg}: The initial value of the trust region radius for the non-gradient optimizer is 0.1
    \end{itemize}
\end{itemize}
\subsection{Results and Observations}
We conducted our experiments on H$_2$ and HeH$^+$, varying the angle \( N \) from -15 degrees to 15 degrees. For each degree, we recorded the estimated energy and the number of iterations required by the optimization algorithm. To evaluate the performance of the Pulse level VQE in the context of a molecule chemistry task. The primary performance metric used in our experiments is the accuracy of the ground state energy, iteration of the optimizer and their deviations. This accuracy is determined by comparing the estimated energy obtained from the Pulse level VQE with the exact ground state energy calculated using the Full Configuration Interaction (FCI) method.
\begin{equation}
\text{Accuracy} = 1 - \left| \frac{E_{\text{VQE}} - E_{\text{FCI}}}{E_{\text{FCI}}} \right|
\end{equation}
where $E_{\text{VQE}}$ is the ground state energy estimated by the Pulse VQE method, and $E_{\text{FCI}}$ is the exact ground state energy obtained using the Full Configuration Interaction (FCI) method.
\begin{equation}
\text{Accuracy Deviation}(N) = \text{Accuracy}(N) - \text{Accuracy}(N=0)
\end{equation}
This equation defines the accuracy deviation at different angles N. Accuracy deviation is the difference between the accuracy at angle N and the accuracy at N=0.
\begin{equation}
\text{Iteration Deviation}(N) = \frac{\text{Iteration}(N) - \text{Iteration}(N=0)}{\text{Iteration}(N=0)}
\end{equation}
This equation defines the iteration deviation at different angles N. Iteration deviation is the relative change in the number of iterations at angle N compared to the number of iterations at N=0.
\begin{figure*}[!t]
\centering
\includegraphics[width=\textwidth]{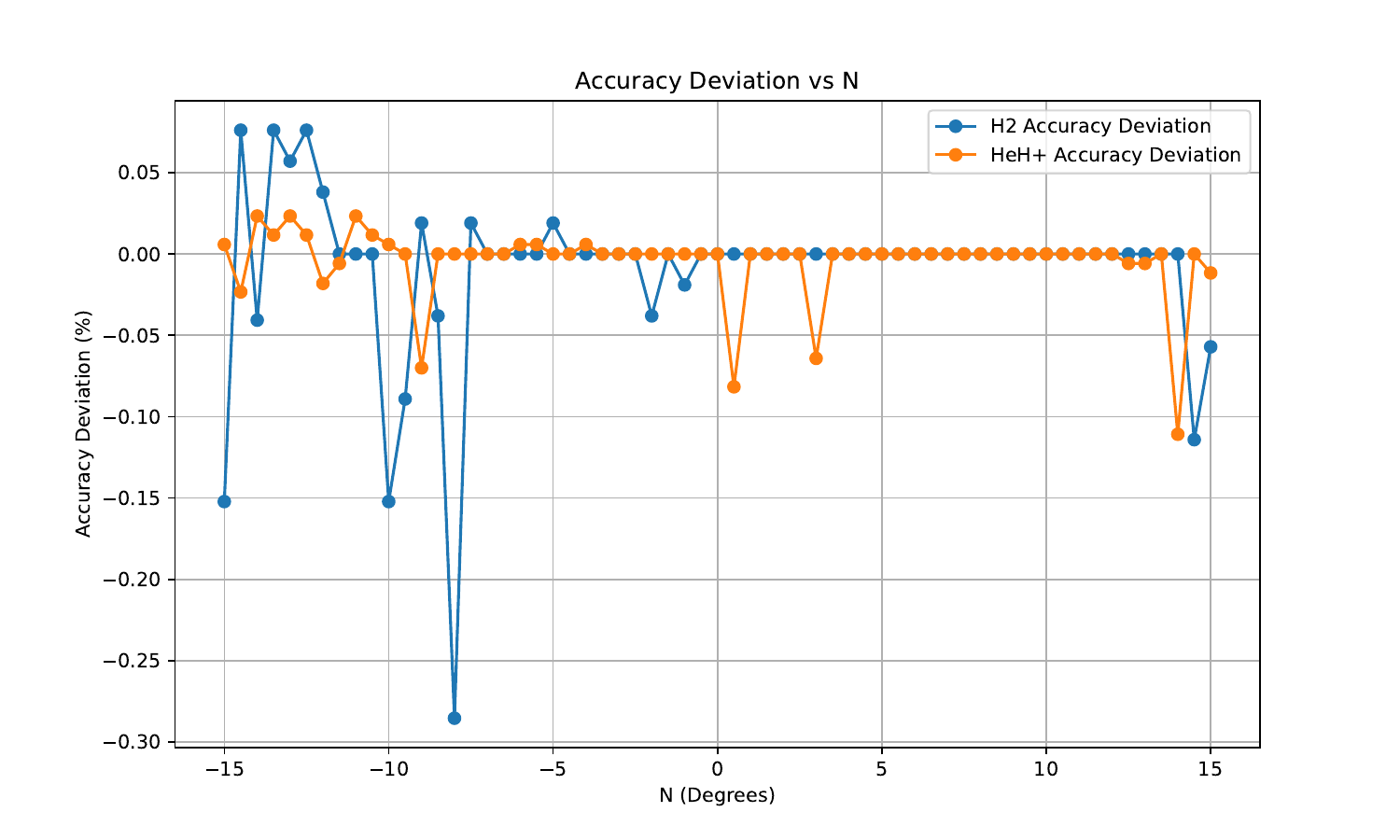}
\caption{The plot shows the accuracy deviation for the molecules H$_2$ and HeH$^+$ as a function of the rotation angle \(N\) (in degrees). The results indicate that both molecules exhibit robustness and performance consistency of the VQE algorithm under different rotation errors. }
\label{fig:accuracy_deviation_vs_N}
\end{figure*}
\begin{figure*}[!t]
\centering
\includegraphics[width=\textwidth]{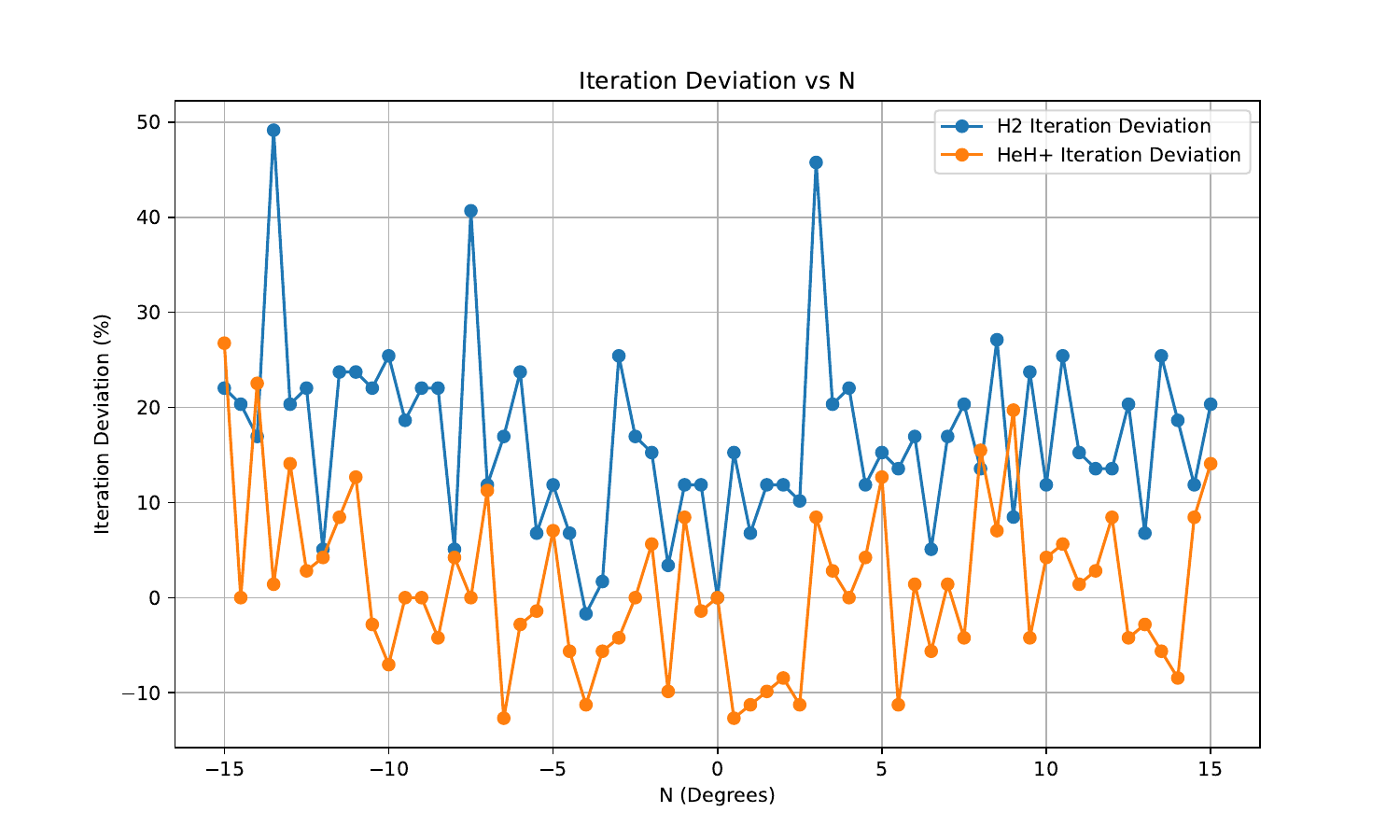}
\caption{plot shows the iteration deviation for the molecules H$_2$ and HeH$^+$ as a function of the rotation angle \(N\) (in degrees). In contract with accuracy deviation, it shows less robustness and performance consistency}
\label{fig:iteration_deviation_vs_N}
\end{figure*}

We provide an abstract of our experimental data in Table \ref{table:results_h2} and Table \ref{table:results_HeH}, which corresponds to \(N\) ranging from -10 degrees to 10 degrees. These data points represent a simplified overview, with significant figures rounded appropriately. This abstract includes a subset of the full data, specifically the integer values from -10 to 10, listed sequentially from -10 to 10. Figure \ref{fig:accuracy_deviation_vs_N} and Figure \ref{fig:iteration_deviation_vs_N} correspond to \(N\) ranging from -15 degrees to 15 degrees, with a step size of 0.5 degrees, representing the complete dataset of our experiment.
Based on the provided data, we can draw the following observations:
\subsubsection{Accuracy Performance}
\begin{table}[ht]
\centering
\caption{Accuracy and Deviations for H\(_2\) and HeH\(^+\)}
\begin{tabular}{|c|c|c|c|}
\hline
Molecule & Accuracy at \( N = 0 \) & Average Accuracy & Max Deviation (\%)\\ \hline
H\(_2\) & 99.70575\% & 99.69582\% & -0.285403 \\ \hline
HeH\(^+\) & 99.964037\% & 99.959724\% & -0.110828 \\ \hline
\end{tabular}
\label{table:accuracy_summary}
\end{table}
These observations showed by Figure \ref{fig:accuracy_deviation_vs_N} and Table \ref{table:accuracy_summary} indicate that the calculated energy values are minimally affected by over-rotation and under-rotation errors in the measurement process as the accuracy remains consistently high and deviation remains consistently low. The results demonstrate high stability and robustness against noise.
\subsubsection{Iteration Performance}
\begin{table}[!t]
\centering
\caption{Correlation Coefficients between Iteration Deviation and Absolute Value of \( N \)}
\label{table:correlation}
\begin{tabular}{ccc}
\hline
Molecule & Positive Correlation & Negative Correlation \\
\hline
\(\text{H}_2\) & 0.0487 & 0.4661 \\
\(\text{HeH}^+\) & 0.3641 & 0.4971 \\
\hline
\end{tabular}
\smallskip
\begin{tablenotes}
\item \textit{Note:} Positive Correlation refers to the correlation coefficient calculated using positive values of \( N \), while Negative Correlation refers to the correlation coefficient calculated using the absolute values of negative \( N \) values.
\end{tablenotes}
\end{table}
\begin{table}[!t]
\centering
\caption{Mean, Standard Deviation, and Maximum Deviation of Iteration Deviation}
\label{table:deviation}
\begin{tabular}{cccc}
\hline
Molecule & Mean & Std Dev & Max Deviation \\
\hline
\(\text{H}_2\) & 0.1675 & 0.0960 & 49.15\% \\
\(\text{HeH}^+\) & 0.0129 & 0.0883 & 26.76\% \\
\hline
\end{tabular}
\end{table}
The correlation coefficients and statistical measures of iteration deviation provide insights into the behavior of our VQE algorithm under varying rotation errors.

As shown in Table \ref{table:correlation}, both the mean and correlation coefficients for iteration deviation are greater than 0. This indicates that while the increase in the rotation angle \( \lvert N \rvert \) does not monotonically increase the iteration count, there is a general trend towards higher iteration counts with increasing \( \lvert N \rvert \).

Table \ref{table:deviation} summarizes the mean, standard deviation, and maximum deviation of iteration deviation. These results indicate that while the VQE algorithm shows resilience to rotation errors in terms of accuracy, the iteration deviation exhibits more variability.

The variability in iteration deviation implies that increasing the rotation error can either increase or decrease the number of iterations required, potentially leading to significant additional computational resource consumption. In extreme cases, this deviation can reach up to 49.15\% for \(\text{H}_2\) and 26.76\% for \(\text{HeH}^+\), which could substantially impact the performance and efficiency of VQE implementations.
\begin{table*}[!t]
\centering
\renewcommand{\arraystretch}{0.5} 
\setlength{\tabcolsep}{0.5pt} 
\caption{Experimental Results for H$_2$}
\label{table:results_h2}
\begin{tabular}{cccccc}
\hline
 \textbf{Energy} & \textbf{Iterations} & \textbf{Iteration Deviation} & \textbf{Accuracy} & \textbf{Accuracy Deviation} \\
\hline
 -1.848982954 & 74 & 25.42\% & 99.55354\% & -0.152215\% \\
 -1.852163385 & 72 & 22.03\% & 99.72478\% & 0.019027\% \\
 -1.846509285 & 62 & 5.08\% & 99.42035\% & -0.285403\% \\
 -1.851810004 & 66 & 11.86\% & 99.70575\% & 0.000000\% \\
-1.851810004 & 73 & 23.73\% & 99.70575\% & 0.000000\% \\
 -1.852163385 & 66 & 11.86\% & 99.72478\% & 0.019027\% \\
 -1.851810004 & 58 & -1.69\% & 99.70575\% & 0.000000\% \\
 -1.851810004 & 74 & 25.42\% & 99.70575\% & 0.000000\% \\
-1.851810004 & 68 & 15.25\% & 99.70575\% & 0.000000\% \\
 -1.851810004 & 66 & 11.86\% & 99.68672\% & -0.019027\% \\
 -1.851810004 & 59 & 0.00\% & 99.70575\% & 0.000000\% \\
 -1.851810004 & 63 & 6.78\% & 99.70575\% & 0.000000\% \\
-1.851810004 & 66 & 11.86\% & 99.70575\% & 0.000000\% \\
 -1.851810004 & 86 & 45.76\% & 99.70575\% & 0.000000\% \\
 -1.851810004 & 72 & 22.03\% & 99.70575\% & 0.000000\% \\
 -1.851810004 & 68 & 15.25\% & 99.70575\% & 0.000000\% \\
 -1.851810004 & 69 & 16.95\% & 99.70575\% & 0.000000\% \\
 -1.851810004 & 69 & 16.95\% & 99.70575\% & 0.000000\% \\
-1.851810004 & 67 & 13.56\% & 99.70575\% & 0.000000\% \\
 -1.851810004 & 64 & 8.47\% & 99.70575\% & 0.000000\% \\
 -1.851810004 & 66 & 11.86\% & 99.70575\% & 0.000000\% \\
\hline
\end{tabular}
\end{table*}
\begin{table*}[!t]
\centering
\renewcommand{\arraystretch}{0.5} 
\setlength{\tabcolsep}{0.5pt} 
\caption{Experimental Results for HeH$^+$}
\label{table:results_HeH}
\begin{tabular}{ccccccc}
\hline
\textbf{Energy} & \textbf{Iterations} & \textbf{Iteration Deviation} & \textbf{Accuracy} & \textbf{Accuracy Deviation} \\
\hline
 -3.92111381 & 66 & -7.04\% &99.96987\% & 0.005833\% \\
 -3.918139537 & 71 & 0.00\% &99.89404\% & -0.069996\% \\
 -3.920885005 & 74 & 4.23\% &99.96403\% & 0.000000\% \\
-3.92188008 & 79 & 11.27\% &99.96403\% & 0.000000\% \\
 -3.921113813 & 69 & -2.82\% &99.96987\% & 0.005834\% \\
 -3.920885005 & 76 & 7.04\% &99.96403\% & 0.000000\% \\
 -3.92111381 & 63 & -11.27\% &99.96987\% & 0.005833\% \\
 -3.920885005 & 68 & -4.23\% &99.96403\% & 0.000000\% \\
 -3.920885005 & 75 & 5.63\% &99.96403\% & 0.000000\% \\
 -3.920885005 & 77 & 8.45\% &99.96403\% & 0.000000\% \\
 -3.920885005 & 71 & 0.00\% &99.96403\% &	0.000000\%	\\
 -3.920885005 & 63 & -11.27\% &99.96403\% & 0.000000\% \\
 -3.920885005 & 65 & -8.45\% &99.96403\% & 0.000000\% \\
 -3.918368231 & 77 & 8.45\% &99.89987\% & -0.064166\% \\
 -3.920885005 & 71 & 0.00\% &99.96403\% & 0.000000\% \\
 -3.920885005 & 80 & 12.68\% &99.96403\% & 0.000000\% \\
 -3.920885005 & 72 & 1.41\% &99.96403\% & 0.000000\% \\
 -3.920885005 & 72 & 1.41\% &99.96403\% & 0.000000\% \\
 -3.920885005 & 82 & 15.49\% &99.96403\% & 0.000000\% \\
 -3.920885005 & 85 & 19.72\% &99.96403\% & 0.000000\% \\
 -3.920885005 & 74 & 4.23\% &99.96403\% & 0.000000\% \\
\hline
\end{tabular}
\end{table*}
\section{Related Work}
NAPA (Native-pulse Ansatz for Variational Quantum Algorithms) is a framework designed to enhance variational quantum algorithms (VQAs) in the Noisy Intermediate Scale Quantum (NISQ) era \cite{liang2024napa}. NAPA aims to exploit VQA advantages by directly using parametric pulses to build ansatz instead of using gates. These native pulses are natively supported on NISQ computers. NAPA generates native-pulse ansatz with the amplitudes and frequencies of pulses as trainable parameters and updates these parameters progressively during training. Their experiments demonstrate this method's ability to decrease latency and achieve high accuracy in VQE tasks.

Previous works~\cite{egger2023pulse,liang2024napa,liang2023towards,meirom2023pansatz, sherbert2024parameterization, liang2023hybrid, meitei2021gate} have primarily focused on the trainable layers of VQEs, specifically on the pulse-level circuits of the ansatz part. For instance, studies have explored various strategies to optimize the parameters of quantum circuits to improve performance and reduce noise susceptibility. These efforts have significantly contributed to our understanding of how to effectively utilize NISQ devices for complex quantum tasks.

However, our work shifts the focus towards the observable aspects of VQEs. We aim to evaluate the quantum errors in measurement steps and investigate to what extent and in what ways these errors influence VQE tasks. By examining the measurement steps, we provide a more detailed analysis of the quantum error impact, which has been largely overlooked in previous studies. This detailed investigation into the observable component complements existing research by offering insights into how measurement errors can affect the overall accuracy and efficiency of VQEs.

By addressing this gap, our research contributes to a more comprehensive understanding of VQE performance in the NISQ era, highlighting the importance of considering both the trainable layers and the measurement steps in quantum algorithm design and implementation.
\section{Discussion}
In this work, we manually introduced over-rotation and under-rotation errors in the observables to evaluate the effects of these quantum errors on the final results of VQEs, focusing on quantum chemistry tasks. Specifically, we studied whether these errors in the rotation pulses could influence the estimated ground state energy and whether they affected the optimizer's performance, potentially requiring more iterations and thus increasing computational costs.

We conducted experiments to calculate the ground state energy of H$_2$ and HeH$^+$ in the presence of rotation errors, recording both the accuracy and the number of iterations required by the optimizer. Our findings indicate that the pulse-level VQE algorithm demonstrates resilience to these quantum errors in terms of accuracy, which remains consistently high. However, the iteration deviation suggests that the computational cost may still be impacted by these errors.

Our research presents several key contributions:

\begin{itemize}
    \item We proposed a novel approach to introduce and evaluate rotation pulse errors in the measurement steps of VQEs. This detailed examination fills a gap in previous works, which often overlooked the impact of noise in observables, potentially leading to overclaimed results.
    \item By providing a specialized discussion on the observable errors, our work supports the design and implementation of pulse-level circuits and algorithms. This includes insights into calibration costs and optimization costs, enabling more informed decisions and improvements.
    \item Given the current state of quantum computers, which are highly susceptible to environmental influences, our work has significant practical implications. Calibration of pulses is often required frequently due to shifts in the system's Hamiltonian, consuming considerable time and computational resources. Our findings suggest that within a certain range, the inaccuracy in the calibration of measurement rotation pulses does not significantly affect the accuracy of the calculations. Therefore, it may be feasible to reduce the frequency of pulse calibrations, thereby saving substantial time and resources. However, this trade-off might lead to increased computational costs due to higher iteration counts, which must be carefully balanced.
    \item An interesting observation from our results is that HeH$^+$, despite having more complex Pauli terms compared to H$_2$, exhibited higher accuracy, smaller maximum accuracy deviation, and lower mean, standard deviation, and maximum deviation in iteration counts. This suggests a stronger resistance to noise for HeH$^+$ calculations. It raises a valuable question: does having more complex Pauli terms enhance noise resistance? This intriguing finding warrants further investigation.
\end{itemize}

In summary, our study highlights the robustness of the pulse-level VQE algorithm in terms of accuracy in the presence of quantum errors in measurement steps and provides valuable insights into optimizing quantum algorithm implementations. By potentially reducing the need for frequent calibrations of measurement rotation pulses, our approach can lead to more efficient use of quantum computing resources. However, this must be balanced against the possibility of increased computational costs. Furthermore, the observed greater noise resistance in HeH$^+$ calculations opens up new avenues for research into the relationship between Pauli term complexity and noise resistance.

\clearpage
\bibliographystyle{ieeetr}
\bibliography{references}

\end{document}